\documentclass[12pt,a4paper]{article}
\usepackage{graphicx}
\usepackage[T1]{fontenc}
\usepackage[utf8]{inputenc}
\usepackage{textcomp}
\usepackage[sc,osf]{mathpazo}
\usepackage{a4wide}  
\usepackage{latexsym,amsthm,amsfonts,amsmath,mathrsfs,amssymb}
\usepackage{dsfont}
\usepackage{accents}
\usepackage[nosort]{cite}
\usepackage{booktabs} 
\usepackage[unicode,implicit]{hyperref}
\hypersetup{%
  pdftitle    = {Magnetic charges and Wald entropy}
  pdfkeywords = {Black holes, thermodynamics, Noether charge, Wald entropy,
    first law, magnetic charge, electric-magnetic duality},
  pdfauthor   = {Tom\'as Ort\'{\i}n and David Pere\~n\'{\i}guez},
  plainpages  = true,
  colorlinks  = true,
  citecolor   = blue,
  urlcolor    = red,
  linkcolor   = black
}
\newcommand{\hepth}[1]{{\tt
\href{http://www.arXiv.org/abs/hep-th/#1}{hep-th/#1}}}
\newcommand{\grqc}[1]{{\tt
\href{http://www.arXiv.org/abs/gr-qc/#1}{gr-qc/#1}}}

\newcommand{\arxiv}[1]{{\tt arXiv:\href{http://www.arXiv.org/abs/#1}{#1}}}
\newcommand{\doi}[1]{{\tt DOI:\href{http://dx.doi.org/#1}{#1}}}

\makeatletter
\@addtoreset{equation}{section}
\makeatother

\begin{document}

\begin{flushright}
\small
IFT-UAM/CSIC-22-40\\
July 26\textsuperscript{th}, 2022\\
\normalsize
\end{flushright}

\vspace{1cm}

\begin{center}

  {\Large {\bf Magnetic charges and Wald entropy}}

\vspace{1.5cm}

\renewcommand{\thefootnote}{\alph{footnote}}

{\sl\large Tom\'as Ort\'{\i}n}\footnote{Email: {\tt  tomas.ortin[at]csic.es}}
{\sl\large  and David Pere\~n\'{\i}guez,}\footnote{Email: {\tt tomas.ortin[at]csic.es}}

\setcounter{footnote}{0}
\renewcommand{\thefootnote}{\arabic{footnote}}
\vspace{1cm}

{\it Instituto de F\'{\i}sica Te\'orica UAM/CSIC\\
C/ Nicol\'as Cabrera, 13--15,  C.U.~Cantoblanco, E-28049 Madrid, Spain}

\vspace{1cm}


{\bf Abstract}
\end{center}
\begin{quotation}
  {\small Using Wald's formalism, we study the thermodynamics (first laws and
    Smarr formulae) of asymptotically-flat black holes, rings etc. in a
    higher-dimensional higher-rank generalization of the Einstein-Maxwell
    theory. We show how to deal with the electric and magnetic charges of the
    objects and how the electric-magnetic duality properties of the theory are
    realized in their thermodynamics.}
\end{quotation}

\newpage
\pagestyle{plain}



\section*{Introduction}

Over most of the past century, electric-magnetic duality has been one of the
main research topics in Theoretical Physics. While it arises as a symmetry of
the source-free equations of motion of electromagnetism in 4 dimensions, it
has been extended and generalized in many directions. Of particular interest
for us is the generalization to higher-rank fields in higher dimensions and
the inclusion of localized sources. The former is quite natural in the context
of brane physics and the second, which goes back to Dirac's discovery of the
magnetic monopole \cite{Dirac:1931kp}, arises naturally when one considers
classical point-like or brane-like sources: the rotations between electric and
magnetic fields must be in one-to-one correspondence with rotations of
electric and magnetic sources and their charges.\footnote{See, for instance,
  Refs.~\cite{Duff:1994an,Ortin:2015hya} and references therein.}

In $d$ dimensions, black holes and black branes can carry some of the electric
and magnetic charges associated to $(p+1)$-form fields: as a general rule,
black $p$-branes (where $p=0$ corresponds to black holes) can carry the
electric charges of $(p+1)$-form potentials and the magnetic charges of
$(\tilde{p}+1)$-form potentials, with $\tilde{p}=d-p-4$ and
$\tilde{\tilde{p}}=p$. When $\tilde{p}=p$, they can carry electric and
magnetic charges of the same $(p+1)$-form field. 

The dynamics of black $p$-branes should exhibit the same electric-magnetic
duality properties as the theory whose equations of motion they solve. In
particular, the laws of thermodynamics \cite{Bardeen:1973gs} should exhibit
those properties. Thus, in $d=4$, where black holes can carry electric and
magnetic charges of the same electromagnetic field, the first law of
black-hole mechanics must include work terms for the variations of both kinds
of charges. The conjugate thermodynamical potentials are the electrostatic and
magnetostatic potentials evaluated on the black-hole horizon. If the theory
has electric-magnetic duality, then we expect the first law of black-hole
mechanics in 4 dimensions and the Smarr formulae \cite{Smarr:1972kt} to be
invariant under the simultaneous rotations of the (variations of the) electric
and magnetic charges and of the conjugate thermodynamical potentials. The
invariance of the Smarr formula for axion-dilaton black holes was recently
proven in Ref.~\cite{Mitsios:2021zrn} using Wald's formalism
\cite{Lee:1990nz,Wald:1993nt,Iyer:1994ys} and the generalized Komar formula
\cite{Komar:1958wp} constructed in
Refs.~\cite{Liberati:2015xcp,Ortin:2021ade}. The proof can be easily
generalized to other theories with scalars and vectors minimally coupled to
gravity with electric-magnetic dualities \cite{Gaillard:1981rj} like most
ungauged supergravity theories. However, a similar proof for the first law
using Wald's formalism is not yet available.

Actually, the presence of magnetic work terms and scalar work terms in the
first law of 4-dimensional black hole mechanics was found by other methods and
is well known \cite{Gibbons:1996af}. In the notation of that reference, it
takes this form:

\begin{equation}
  \delta M
  =
  \frac{\kappa \delta A}{8\pi G_{N}^{(4)}}+\Omega\delta J
  +\psi^{\Lambda}\delta q_{\Lambda}+\chi_{\Lambda}\delta p^{\Lambda}
  -G_{ab}(\phi_{\infty})\Sigma^{a}\delta \phi^{b}_{\infty}\,,
\end{equation}

\noindent
where the $q_{\Lambda}$s and $p_{\Lambda}$s are, respectively, electric and
magnetic charges with respect to the vector field $A^{\Lambda}$ and
$\psi^{\Lambda}$ and $\chi_{\Lambda}$ are the electrostatic and magnetostatic
potentials evaluated on the horizon, $G_{ab}(\phi_{\infty})$ is the metric of
the scalar manifold evaluated at infinity, $\Sigma^{a}$ the scalar charges and
$\phi^{a}_{\infty}$ the vales of the scalars at infinity (moduli).

The invariance of this formula under electric-magnetic duality transformations
presents several problems because, on general grounds \cite{Gaillard:1981rj},
the terms involving electric and magnetic charges should be combined in a
manifestly symplectic-invariant expression, which is not the case. The term
involving the scalar charges is manifestly invariant. However, there is no
good definition of the scalar charge as a conserved charge and, while this
does not invalidate the result, it obscures its meaning.

In this paper we want to study the electric-magnetic duality properties of the
first law of black hole mechanics for asymptotically-flat black $p$-branes
coupled to higher-rank form potentials in $d$ dimensions\footnote{For $p=0$
  these are black holes, for $p=1$ black rings etc. Standard black $p$-branes
  are generically not asymptotically flat because they either extend to
  infinity or they are wrapped around compact dimensions and, in both cases,
  they would only be asymptotically flat in the transverse, non-compact
  dimensions. They have to be studied separately. Work in this direction is
  already in progress \cite{kn:MOPZ}.} using Wald's formalism, leaving the
problem of understanding the term with scalar charges for later work. In
previous work \cite{Elgood:2020svt,Elgood:2020mdx,Elgood:2020nls} we showed
how to prove the first law in presence of matter fields by correctly taking
into account the interplay between diffeomorphisms and gauge
transformations. However, since there is only one gauge transformation per
gauge field, we were unable to recover the work terms proportional to the
variations of the magnetic charges or the moduli, even though the same methods
correctly give the terms proportional to magnetic charges in the Smarr formula
Refs.~\cite{Liberati:2015xcp,Ortin:2021ade,Mitsios:2021zrn}. In this paper we
are going to show how those terms arise in a more careful calculation and how
they do it with the appropriate sign to have electric-magnetic duality
invariance.

The theory we are going to consider is the straightforward generalization of
Einstein-Maxwell to higher dimensions and higher-rank form potentials. Thus,
it differs from the one considered in Ref.~\cite{Compere:2007vx}\footnote{See
  also Ref.~\cite{Copsey:2005se}.}  by the absence of scalar fields, which we
plan to study in future work.  In Ref.~\cite{Compere:2007vx} only electric
charges and electrically-charged black branes were considered and, in this
work, we are going to show how to include the magnetic ones. We will not
constrain the dimension of the horizon and, therefore, we consider,
simultaneously, black $p$-branes which are electrically charged with respect
to the $(p+1)$-form field and black $\tilde{p}$-branes which are magnetically
charged with respect to it or electrically with respect to the dual
$(\tilde{p}+1)$-form field, although we are always going to use the
formulation in which the $(p+1)$-form potential appears.\footnote{In
  Ref.~\cite{Meessen:2022hcg} we have considered an example in which both the
  fundamental and the dual potential occur in the action. There are two gauge
  symmetries and the inclusion of magnetic charges is straightforward. These
  democratic formulations are often very complicated and here we are not going
  to use them even though in this case they would be much simpler to find.}

This paper is organized as follows: in Section~\ref{sec-emduality} we
introduce the theories we are going to consider and we are going to explain
how electric-magnetic duality is realized in them. In
Section~\ref{sec-conservedcharges} we are going to define the conserved
charges of the theory: those associated to the gauge symmetries and the
magnetic ones, whose nature is topological. In Section~\ref{sec-zerothlaws} we
are going to prove the restricted form of the generalized zeroth laws which we
will use in Section~\ref{sec-Smarrformulae} to find the Smarr formulae and in
Section~\ref{sec-firstlaw} to prove the first law. We end by discussing the
results obtained and proposing new directions of research in
Section~\ref{sec-discussion}.

\section{Electric-magnetic duality and $(p+1)$-forms}
\label{sec-emduality}

We are going to consider the generalization of the $d$-dimensional
Einstein-Maxwell theory in which the Maxwell 1-form field is replaced by a
$(p+1)$-form $A$\footnote{In our notation $p$ is the dimension of the objects
  that couple to these potentials, namely $p$-branes with $(p+1)$-dimensional
  worldvolumes. This notation differs from the one used in
  Ref.~\cite{Compere:2007vx}, but it is the most natural one in brane
  physics.}

\begin{equation}
  A
  =
  \frac{1}{(p+1)!}A_{\mu_{1}\cdots \mu_{p+1}}dx^{\mu_{1}}\wedge \cdots \wedge
  dx^{\mu_{p+1}}\,,  
\end{equation}

\noindent
whose $(p+2)$-form field strength $F$\footnote{With our normalization, the
  components of $F$ are defined by
  \begin{equation}
    F
    =
    \frac{1}{(p+2)!}F_{\mu_{1}\cdots \mu_{p+2}}dx^{\mu_{1}}\wedge \cdots \wedge
    dx^{\mu_{p+2}}\,,
    \end{equation}
    so
    \begin{equation}
      F_{\mu_{1}\cdots \mu_{p+2}}
      =
      (p+2)\partial_{[\mu_{1}}A_{\mu_{2}\cdots \mu_{p+2}]}\,.
    \end{equation}
} 

\begin{equation}
  F
  \equiv
  dA
  =
  \frac{1}{(p+1)!}\partial_{\mu_{1}}A_{\mu_{2}\cdots \mu_{p+2}}
  dx^{\mu_{1}} \wedge \cdots \wedge dx^{\mu_{p+2}}\,,
\end{equation}

\noindent
is invariant under gauge transformations

\begin{equation}
  \label{eq:gaugetrans}
\delta_{\chi}A = d\chi\,,  
\end{equation}

\noindent
where $\chi$ is an arbitrary $p$-form.

The components of the dual $(\tilde{p}+2)$-form $\star F$
($\tilde{p}\equiv d-p-4$) are given by

\begin{equation}
  (\star F)_{\mu_{1}\cdots \mu_{\tilde{p}+2}}
  \equiv
  \frac{1}{(p+2)!\sqrt{|g|}}
  \varepsilon_{\mu_{1}\cdots \mu_{\tilde{p}+2}\nu_{1}\cdots \nu_{p+2}}
  F^{\nu_{1}\cdots \nu_{p+2}}\,,
\end{equation}

\noindent
and

\begin{equation}
  \star^{2} F = \sigma^{2} F\,,
  \,\,\,\,\,\,\,
  \text{with}
  \,\,\,\,\,\,\,
  \sigma^{2} = (-1)^{(d+1)(p+1)}\,.
\end{equation}

We will call the dual form $G$

\begin{equation}
G \equiv \star F\,.  
\end{equation}

\noindent
$G$ and $F$ are forms of the same rank when $p=\tilde{p}$, which happens when
$d=2(p+2)$. Real (anti-) self-duality requires $\sigma^{2}=+1$ and only
$(p+1)$-form fields with $p$ odd can have this property.

We choose  the Vielbein

\begin{equation}
  e^{a}
  =
  e^{a}{}_{\mu}dx^{\mu}\,,
\end{equation}

\noindent
as the gravitational field. The Levi-Civita spin connection
$\omega^{ab}=-\omega^{ba}$ is defined through the first Cartan structure
equation

\begin{equation}
  \mathcal{D} e^{a}
  =
  de^{a} -\omega^{a}{}_{b}\wedge e^{b}
  =
  0\,,
\end{equation}

\noindent
and the curvature 2-form is

\begin{equation}
  R^{ab}
  =
  d\omega^{ab} -\omega^{a}{}_{c}\wedge \omega^{cb}\,.
\end{equation}

\noindent
We will also use the total covariant derivative $\nabla$. It satisfies the
Vielbein postulate

\begin{equation}
  \nabla_{\mu}e^{a}{}_{\nu}
  -\omega_{\mu}{}^{a}{}_{b}e^{b}{}_{\nu}
  -\Gamma_{\mu\nu}{}^{\rho}e^{a}{}_{\rho}
  =
  0\,,
\end{equation}

\noindent
that relates the components of the spin connection $\omega_{\mu}{}^{ab}$ to
those of the affine connection $\Gamma_{\mu\nu}{}^{\rho}$ which are given by
the Christoffel symbols

\begin{equation}
  \Gamma_{\mu\nu}{}^{\rho}
  =
  \tfrac{1}{2}g^{\rho\sigma}\left\{\partial_{\mu}g_{\nu\sigma}
  +\partial_{\nu}g_{\mu\sigma}-\partial_{\sigma}g_{\mu\nu}\right\}\,.
\end{equation}

In terms of these variables and objects, the action we want to consider is

\begin{equation}
  \label{eq:action}
  \begin{aligned}
    S[e^{a},A]
    & =
    \frac{1}{16\pi G_{N}^{(d)}} \int \left\{(-1)^{d-1}\star
      (e^{a}\wedge e^{b})\wedge R_{ab} +\frac{(-1)^{d(p-1)}}{2}F\wedge \star F
    \right\}
    \\
    & \\
    & \equiv
    \int \mathbf{L}\,.
  \end{aligned}
\end{equation}

\noindent
Since we are mainly interested in the electric-magnetic duality properties of
the first law and Smarr formulae, we are not including the dilaton field that
usually couples to these forms in supergravity/superstring theories. A more
general study including general couplings to scalars will be made elsewhere.

Under a general variation of the fields

\begin{equation}
  \label{eq:variationaction}
  \delta S
  =
  \int \left\{\mathbf{E}_{a}\wedge \delta e^{a} +\mathbf{E}\wedge \delta A
  +d\mathbf{\Theta}(e^{a},A,\delta e^{a},\delta A)\right\}\,,
\end{equation}

\noindent
where the equations of motion $\mathbf{E}_{a}$ (Einstein) and $\mathbf{E}$
(Maxwell) and the symplectic potential $(d-1)$-form
$\mathbf{\Theta}(e^{a},A,\delta e^{a},\delta A)$ are given by\footnote{In
  order to simplify the expressions, we suppress the global factors of
  $\left(16\pi G_{N}^{(d)}\right)^{-1}$. We will restore them in the final
  results.}

\begin{subequations}
  \begin{align}
    \label{eq:Ea}
  \mathbf{E}_{a}
  & =
    \imath_{a}\star (e^{c}\wedge e^{d})\wedge R_{cd}
    +\tfrac{(-1)^{dp}}{2}\left(\imath_{a}F\wedge G+(-1)^{p+1}F\wedge \imath_{a}G\right)\,,
\\
    & \nonumber \\
    \label{eq:E}
    \mathbf{E}
    & =
      -dG\,,
    \\
    & \nonumber \\
      \mathbf{\Theta}(e^{a},A,\delta e^{a},\delta A)
  & =
     -\star (e^{a}\wedge e^{b})\wedge \delta \omega_{ab}
    +G\wedge \delta A\,, 
  \end{align}
\end{subequations}

\noindent
where $\imath_{c}$ stands for $\imath_{e_{c}}$, \textit{i.e.}~the interior
product with the vector field $e_{c}=e_{c}{}^{\mu}\partial_{\mu}$.

This set of equations can be enlarged with the Bianchi identity

\begin{equation}
  \label{eq:B}                  
\mathbf{B} \equiv -dF\,.  
\end{equation}

\noindent
In order to explore the invariance of the enlarged set of equations of motion
under electric-magnetic duality transformations, it is convenient to define
the vector of field strengths

\begin{equation}
\mathcal{F}
\equiv
\left(
  \begin{array}{c}
  F \\ G \\  
  \end{array}
\right)\,, 
\end{equation}

\noindent
in terms of which the equations take the form

\begin{subequations}
  \begin{align}
    \label{eq:Eadual}
  \mathbf{E}_{a}
  & =
    \imath_{a}\star (e^{c}\wedge e^{d})\wedge R_{cd}
    -\tfrac{(-1)^{p(d+1)}}{2}\mathcal{F}^{T}\Omega\,
    \imath_{a}\mathcal{F}\,,
\\
    & \nonumber \\
    \label{eq:EB}
    \left(
    \begin{array}{c}
\mathbf{B} \\ \mathbf{E} 
    \end{array}
\right)
    & =
      -d\mathcal{F}\,,
  \end{align}
\end{subequations}

\noindent
where we have defined the $2\times 2$ matrix

\begin{equation}
  \Omega
  \equiv
  \left(
    \begin{array}{cc}
     0 & 1 \\ & \\ \sigma^{2} & 0 \\ 
    \end{array}
    \right)\,.
\end{equation}

\noindent
For $\sigma^{2}=+1$, this matrix is the non-diagonal metric of O$(1,1)$ and
for $\sigma^{2}=-1$ it is the ``metric'' of Sp$(2,\mathbb{R})$. It is, then,
evident, that the above system of equations is invariant under linear
transformations of $\mathcal{F}$ and that the groups of invariance are
O$(1,1)$ and Sp$(2,\mathbb{R})$. Of course, these linear transformations only
make sense for $p=\tilde{p}$. However, when this is not the case, the
equations are still formally invariant under the discrete subgroups of
O$(1,1)$ and Sp$(2,\mathbb{R})\sim$SL$(2,\mathbb{R})$ that simply interchange
$F$ and $G$ (up to signs).

However, these are not the duality groups of the theory because the
transformations must respect the self-duality constraint

\begin{equation}
  \star \mathcal{F}
  =
  \Omega \mathcal{F}\,.  
\end{equation}

\noindent
For $\sigma^{2}=+1$, only the interchange of $F$ and $G$ (up to a global sign)
survives, while, for $\sigma^{2}=-1$, it is the whole continuous subgroup
SO$(2)\subset~$SL$(2,\mathbb{R})$ that survives.

Observe that the symplectic potential
$\mathbf{\Theta}(e^{a},A,\delta e^{a},\delta A)$ is not invariant under any
electric-magnetic duality transformations. This is not surprising because the
action is not invariant, either. 

\section{Conserved charges}
\label{sec-conservedcharges}

\subsection{Lorentz charge}
\label{sec-lorentzcharge}

The action Eq.~(\ref{eq:action}) is exactly invariant under local Lorentz
transformations of the Vielbein

\begin{equation}
    \delta_{\sigma}e^{a}
    =
    \sigma^{a}{}_{b}e^{b}\,,  
\end{equation}

\noindent
which induce the following transformation of the spin connection and curvature

\begin{equation}
  \begin{aligned}
    \delta_{\sigma}\omega^{ab}  
    & =
    \mathcal{D}\sigma^{ab}\,,
    \\
    & \\
    \delta_{\sigma}R^{ab}
    & =
    2\sigma^{[a|}{}_{c}R^{c|b]}\,.
  \end{aligned}
\end{equation}

For these particular transformations and upon use of the Noether identity (the
symmetry of the Einstein equations)

\begin{equation}
  \label{eq:LorentzNoetheridentity}
\mathbf{E}^{[a}\wedge e^{b]}=0\,,
\end{equation}

\noindent
we find 

\begin{equation}
  \label{eq:variationactionsigma}
  \delta_{\sigma} S
  =
  \int d \mathbf{J}[\sigma]\,,
  \hspace{1cm}
  \mathbf{J}[\sigma]
  =
  -\star (e^{a}\wedge e^{b})\wedge  \mathcal{D}\sigma_{ab}\,.
\end{equation}

The off-shell invariance of the action for arbitrary parameters $\sigma^{ab}$
and arbitrary integration region imply the closedness of $\mathbf{J}[\sigma]$
and its local exactness

\begin{equation}
  \mathbf{J}[\sigma]
  =
  d\mathbf{Q}[\sigma]\,,
  \hspace{1cm}
  \mathbf{Q}[\sigma]
  =
  \frac{(-1)^{d-1}}{16\pi G_{N}^{(d)}}\star (e^{a}\wedge e^{b})\wedge \sigma_{ab}\,.
\end{equation}

Using this $(d-2)$-form one can construct for on-shell field configurations a
conserved charge for each independent parameter $\sigma^{ab}$ that leaves that
field configuration invariant \cite{Barnich:2001jy}. However, there are no
non-trivial parameters $\sigma^{ab}$ that leave invariant a regular Vielbein
$\sigma^{a}{}_{b}e^{b}=0$ and, therefore, there seems to be no conserved
charges associated to this symmetry. In spite of this, this $(d-2)$-form plays
an important role in black-hole thermodynamics for a particular $\sigma^{ab}$,
as we are going to see and we have already pointed out in
Refs.~\cite{Elgood:2020mdx,Elgood:2020nls}.

\subsection{Electric charge}
\label{sec-electriccharge}

The action Eq.~(\ref{eq:action}) is exactly invariant under the gauge
transformations of the $(p+1)$-form field $A$ Eq.~(\ref{eq:gaugetrans}). For
those particular transformations, and using the Noether identity

\begin{equation}
  \label{eq:MaxwellNoetheridentity}
d\mathbf{E}=0\,,  
\end{equation}

\noindent
the general variation of the action Eq.~(\ref{eq:variationaction}) can be
written in the form

\begin{equation}
  \delta_{\chi}S
  =
  \int d\mathbf{J}[\chi]\,,
  \hspace{1cm}
  \mathbf{J}[\chi]
  =
  (-1)^{d-p+1}\mathbf{E}\wedge \chi +G\wedge d\chi\,.
\end{equation}

\noindent
The off-shell invariance of the action for arbitrary $p$-forms $\chi$ and
integration region imply the closedness of $\mathbf{J}[\chi]$ and its local
exactness

\begin{equation}
  \mathbf{J}[\chi] = d\mathbf{Q}[\chi]\,,
  \hspace{1cm}
  \mathbf{Q}[\chi]
  =
  \frac{(-1)^{d(p-1)}}{16\pi G_{N}^{(d)}} \chi\wedge G\,.
\end{equation}

The $(d-2)$-form $\mathbf{Q}[\chi]$ can be used to define a charge which is
conserved on-shell ($dG=0$) for each independent gauge parameter $\chi$
leaving invariant the field configuration
\cite{Barnich:2001jy,Barnich:2003xg}. The $p$-form gauge parameters that leave
invariant the potential $A$ are the closed ones $d\chi=0$. In a compact
manifold with no boundary, these can be decomposed in a linear combination of
harmonic $p$-forms $h_{i}$ plus an exact $p$-form $de$. Only the harmonic ones
give non-trivial conserved charges when integrated over closed codimension-2
surfaces $\Sigma^{d-2}$

\begin{equation}
  \label{eq:electriccharges}
  Q_{i}
  \equiv
   \frac{(-1)^{d(p+1)}}{16\pi G_{N}^{(d)}} \int_{\Sigma^{d-2}} h_{i}\wedge G\,,
\end{equation}

\noindent
and the addition of exact $p$-forms to $h_{i}$ does not change their values
\cite{Copsey:2005se,Compere:2007vx}. Observe that the sign we have chosen in
this definition is purely conventional.

\subsection{Magnetic charge}
\label{sec-magneticcharge}

Even though there are no more gauge symmetries in our theory, we can define
magnetic charges which are conserved in exactly the same sense as the electric
ones:

\begin{equation}
  \label{eq:magneticcharges}
  P^{m}
  \equiv
  \frac{(-1)^{d(p+1)}}{16\pi G_{N}^{(d)}}
  \int_{\Sigma^{d-2}} \tilde{h}^{m}\wedge F\,,  
\end{equation}

\noindent
where $\tilde{h}^{m}$ is a harmonic $\tilde{p}$-form. We are using a different
set of indices for the magnetic charges since, in general, the number of
harmonic $p$- and $\tilde{p}$-forms need not be the same. It is unclear how
duality rotations or Dirac-like quantization conditions for these charges can
be defined, except for the special case in which $p=\tilde{p}$ and
$\tilde{h}^{i}=h_{i}$. In this case, the charges can also be arranged in
vectors

\begin{equation}
  \label{eq:chargevector}
  \mathcal{Q}_{i}
  \equiv
  \left(
    \begin{array}{c}
      P_{i} \\ Q_{i} \\
    \end{array}
  \right)
  =
  \frac{1}{16\pi G_{N}^{(d)}} \int_{\Sigma^{d-2}} h_{i}\wedge \mathcal{F}\,,
\end{equation}

\noindent
transforming in the same way as $\mathcal{F}$.

Observe that the definition of magnetic charge Eq.~(\ref{eq:magneticcharges})
becomes trivial and gives zero whenever $F=dA$ \textit{globally}. Thus, as
expected, non-vanishing magnetic charges are an exclusive property of certain
non-trivial gauge field configurations.

\subsection{Noether-Wald charge}
\label{sec-NoetherWaldcharge}

The action Eq.~(\ref{eq:action}) is exactly invariant under infinitesimal
diffeomorphisms

\begin{equation}
  \delta x^{\mu} = \xi^{\mu}(x)\,.  
\end{equation}

\noindent
Then, if we consider only the infinitesimal transformations of the fields
$\delta \varphi \equiv \varphi'(x)-\varphi(x)$ the action is invariant up to a
total derivative

\begin{equation}
  \label{eq:deltaxiS}
  \delta_{\xi}S
  =
  -\int d\imath_{\xi}\mathbf{L}\,.
\end{equation}

\noindent
The associated $(d-2)$-form $\mathbf{Q}[\xi]$ is called the Noether-Wald
charge \cite{Wald:1993nt}.

As discussed in
Refs.~\cite{Elgood:2020svt,Elgood:2020mdx,Elgood:2020nls,Mitsios:2021zrn,Meessen:2022hcg},\footnote{A
  different, more mathematically rigorous approach based on the theory of
  principal bundles was followed in Ref.~\cite{Prabhu:2015vua}, but it cannot
  be applied to the $p>0$ gauge transformations considered here or in
  Refs.~\cite{Elgood:2020mdx,Elgood:2020nls,Meessen:2022hcg}.} the
transformation of fields with some gauge freedom under infinitesimal
diffeomorphisms is, generically, of the form

\begin{equation}
  \label{eq:deltaxidef}
  \delta_{\xi} = -\pounds_{\xi}+\delta_{\Lambda_{\xi}}\,,   
\end{equation}

\noindent
where $\pounds_{\xi}$ is the standard Lie derivative with respect to the
vector field $\xi$ and $\delta_{\Lambda_{\xi}}$ is a (``compensating'' or
``induced'') gauge transformation whose parameter $\Lambda_{\xi}$ depends on
$\xi$ and on the fields on which the transformation acts.\footnote{A slightly
  different point of view is that of ``invariance up to gauge
  transformations'', taken in
  Refs.~\cite{Sudarsky:1992ty,Sudarsky:1993kh,Barnich:2007bf,McCormick:2013nkb,Hajian:2013lna,Hajian:2015xlp}.}
  We should write, then, $\Lambda(\xi,e^{a},A)$, but we will use
  $\Lambda_{\xi}$ for simplicity, keeping in mind the dependence on the
  fields.  In general the value of this parameter is only fully determined
  when the diffeomorphism is a symmetry of the whole field configuration. In
  that case we will denote the vector field that generates it by $k$ since, in
  particular, it must be a Killing vector and one can define

\begin{equation}
  \delta_{k} = -\pounds_{k}+\delta_{\Lambda_{k}} \equiv -\mathbb{L}_{k}\,,   
\end{equation}

\noindent
where $\mathbb{L}_{k}$ transforms covariantly under gauge transformations,
hence the name \textit{covariant Lie derivative}. This property (which is not
shared by the standard Lie derivative) has to be checked case by case. It
ensures that the annihilation of all the fields by the transformation
$\delta_{k}$ (or by the operator $\mathbb{L}_{k}$) is a gauge-independent
condition.

In the case of the $(p+1)$-form field $A$, the compensating $p$-form gauge
parameter is given by \cite{Elgood:2020svt,Elgood:2020mdx,Elgood:2020nls}

\begin{equation}
  \label{eq:chixi}
  \chi_{\xi}
  =
  \imath_{\xi}A-P_{\xi}\,,
\end{equation}

\noindent
where the \textit{momentum map} $p$-form $P_{\xi}$ satisfies, for $\xi=k$, the
\textit{momentum map equation}

\begin{equation}
  \label{eq:maxwellmomentummap}
dP_{k}+\imath_{k}F=0\,.  
\end{equation}

\noindent
Then

\begin{equation}
  \delta_{\xi}A
  =
  -\pounds_{\xi}A+\delta_{\chi_{\xi}} A
  =
  -(d\imath_{\xi} +\imath_{\xi}d)A+d\chi_{\xi}
=
-(dP_{\xi}+\imath_{\xi}F)\,,
\end{equation}

\noindent
which is guaranteed to vanish when $\xi=k$ by virtue of the momentum map
equation (\ref{eq:maxwellmomentummap}).

The $(p+2)$-form field strength $F$ is gauge invariant and, upon use of the
Bianchi identity

\begin{equation}
  \delta_{k} F
  =
  -\pounds_{k} F
  =
  -(d\imath_{k} +\imath_{k}d)F
  =
  -d\imath_{k}F\,,
\end{equation}

\noindent
which, yet again, vanishes identically by virtue of the momentum map equation
(\ref{eq:maxwellmomentummap}).

In the case of the Vielbein field, the compensating gauge (Lorentz) parameter
is given by
\cite{Ortin:2002qb,Jacobson:2015uqa,Elgood:2020svt,Elgood:2020mdx,Elgood:2020nls}\footnote{The
  resulting covariant derivative, known as \textit{Lie-Lorentz covariant
    derivative} is a generalization of the spinorial derivative of
  Refs.~\cite{kn:Lich,kn:Kos,kn:Kos2,Hurley:cf,Ortin:2002qb}.}

\begin{equation}
  \label{eq:sigmaxi}
  \sigma_{\xi}{}^{ab}
  =
  \imath_{\xi}\omega^{ab} -P_{\xi}{}^{ab}\,,
\end{equation}

\noindent
where $P_{\xi}{}^{ab}$ is the \textit{Lorentz momentum map} which, for
$\xi=k$, is defined to satisfy the \textit{Lorentz momentum map equation} 

\begin{equation}
  \label{eq:lorentzmomentummap}
  \mathcal{D}P_{k}{}^{ab}+\imath_{k}R^{ab}
  =
  0\,.
\end{equation}

This equation is solved by the \textit{Killing bivector}

\begin{equation}
  P_{k}{}^{ab}
  =
  \nabla^{a}k^{b}
  =
  \nabla^{[a}k^{b]}\,.
\end{equation}

\noindent
As a matter of fact, for this value of the momentum map, the Lorentz momentum
map equation (\ref{eq:lorentzmomentummap}) becomes the integrability condition
of the Killing vector equation. Then, on the Vielbein

\begin{equation}
  \delta_{\xi}e^{a}
  =
  -(d\imath_{\xi} +\imath_{\xi}d)e^{a} +\sigma_{\xi}{}^{a}{}_{b}e^{b}
  =
   \mathcal{D}\xi^{a} +P_{\xi}{}^{a}{}_{b}e^{b}
  =
  -\tfrac{1}{2}\left(\nabla_{\mu}\xi^{a} +\nabla^{a}\xi_{\mu}\right)dx^{\mu}\,,
\end{equation}

\noindent
which vanishes when $\xi=k$ by virtue of the Killing vector equation.

For the spin connection we have

\begin{equation}
  \delta_{\xi}\omega^{ab}
  =
  -(d\imath_{\xi} +\imath_{\xi}d)\omega^{ab} +\mathcal{D}\sigma_{\xi}{}^{ab}
  =
  -\left(\mathcal{D}P_{\xi}{}^{ab}+\imath_{\xi}R^{ab}\right)\,,
\end{equation}

\noindent
that vanishes when $\xi=k$ by virtue of the Lorentz momentum map equation
(\ref{eq:lorentzmomentummap}).

Finally, as a simple exercise, we can consider the transformation of the
curvature. According to the general rule and using the explicit form of the
compensating Lorentz parameter, we get

\begin{equation}
  \begin{aligned}
    \delta_{\xi}R^{ab}
    & =
    -\pounds_{\xi}R^{ab} +2\sigma_{\xi}{}^{[a|}{}_{c}R^{c|b]}
    \\
    & \\
    & =
    -\imath_{\xi}\left(\mathcal{D}R^{ab}+2\omega^{[a|}{}_{c}R^{c|b]}\right)
    -\left(\mathcal{D}\imath_{\xi}R^{ab}+2\omega^{[a|}{}_{c}\imath_{\xi}R^{c|b]}\right)
    \\
    & \\
    & \hspace{.5cm}
    +2\imath_{k}\omega^{[a|}{}_{c}R^{c|b]} -2P_{\xi}{}^{[a|}{}_{c}R^{c|b]}
    \\
    & \\
    & =
    -\mathcal{D}\imath_{\xi}R^{ab} -2P_{\xi}{}^{[a|}{}_{c}R^{c|b]}\,.
  \end{aligned}
\end{equation}

\noindent
When $\xi=k$ we can use the Lorentz momentum map equation and 

\begin{equation}
    \delta_{k}R^{ab}
    =
    \mathcal{D}\mathcal{D}P_{k}{}^{ab} -2P_{k}{}^{[a|}{}_{c}R^{c|b]}\,,
\end{equation}

\noindent
which vanishes identically (Ricci identity).

In order to find the Noether-Wald charge we just have to plug the above
transformations into the general variation of the action
Eq.~(\ref{eq:variationaction})

\begin{equation}
  \label{eq:variationactionxi}
  \delta_{\xi} S
  =
  \int \left\{-\mathbf{E}_{a}\wedge \left(  \mathcal{D}\xi^{a}
      +P_{\xi}{}^{a}{}_{b}e^{b}\right)
    -\mathbf{E}\wedge \left(dP_{\xi}+\imath_{\xi}F \right)
  +d\mathbf{\Theta}(e^{a},A,\delta_{\xi} e^{a},\delta_{\xi} A)\right\}\,,
\end{equation}

\noindent
with

\begin{equation}
  \mathbf{\Theta}(e^{a},A,\delta_{\xi} e^{a},\delta_{\xi} A)
  =
  \star (e^{a}\wedge e^{b})\wedge
  \left(\mathcal{D}P_{\xi}{}^{ab}+\imath_{\xi}R^{ab}\right)
    +G\wedge \left(dP_{\xi}+\imath_{\xi}F \right)\,.
\end{equation}

The term involving $P_{\xi}{}^{ab}$ in Eq.~(\ref{eq:variationactionxi})
vanishes by virtue of the Noether identity associated to local Lorentz
invariance Eq.~(\ref{eq:LorentzNoetheridentity}).  Integrating by parts the
two terms of Eq.~(\ref{eq:variationactionxi}) that involve derivatives, we get

\begin{equation}
  \label{eq:variationactionxi1}
  \delta_{\xi} S
  =
  \int \left\{(-1)^{d-1}\mathcal{D}\mathbf{E}_{a} \xi^{a} 
    -\mathbf{E}\wedge\imath_{\xi}F
    +(-1)^{d-p-1}d\mathbf{E}\wedge P_{\xi}
  +d\mathbf{\Theta}'(e^{a},A,\delta_{\xi} e^{a},\delta_{\xi} A)\right\}\,,
\end{equation}

\noindent
with

\begin{equation}
  \mathbf{\Theta}'(e^{a},A,\delta_{\xi} e^{a},\delta_{\xi} A)
  =
  \mathbf{\Theta}(e^{a},A,\delta_{\xi} e^{a},\delta_{\xi} A)
  +(-1)^{d}\mathbf{E}_{a} \xi^{a} +(-1)^{d-p}\mathbf{E}\wedge P_{\xi}\,.
\end{equation}

\noindent
Using the Noether identities associated to gauge transformations
Eq.~(\ref{eq:MaxwellNoetheridentity}) and diffeomorphisms\footnote{The proof
  of this identity is a trivial generalization of the proof given in
  Ref.~\cite{Elgood:2020svt} for the case $p=0$.}

\begin{equation}
\mathcal{D}\mathbf{E}_{a} \xi^{a} +(-1)^{d}\mathbf{E}\wedge\imath_{\xi}F
=
0\,,
\end{equation}

\noindent
we arrive at

\begin{equation}
  \label{eq:variationactionxi2}
  \delta_{\xi} S
  =
  \int d\mathbf{\Theta}'(e^{a},A,\delta_{\xi} e^{a},\delta_{\xi} A)\,,
\end{equation}

\noindent
which, combined with Eq.~(\ref{eq:deltaxiS}), leads to

\begin{equation}
  \label{eq:Jxidef}
  d\mathbf{J}[\xi]
  =
  0\,,
  \hspace{1cm}
  \mathbf{J}[\xi]
  \equiv
  \mathbf{\Theta}'(e^{a},A,\delta_{\xi} e^{a},\delta_{\xi} A)
  +\imath_{\xi}\mathbf{L}\,.
\end{equation}

\noindent
As usual, this implies the local existence of the $(d-2)$-form
$\mathbf{Q}[\xi]$

\begin{equation}
  \mathbf{J}[\xi]
  =
  d\mathbf{Q}[\xi]\,,
  \hspace{1cm}
  \mathbf{Q}[\xi]
  =
  \frac{1}{16\pi G_{N}^{(d)}}
  \left\{(-1)^{d}\star (e^{a}\wedge e^{b})P_{\xi\, ab}
    -(-1)^{d(p-1)}P_{\xi}\wedge G\right\}\,,
\end{equation}

\noindent
which is a straightforward generalization of the Noether-Wald $(d-2)$-form
obtained in the Einstein-Maxwell case in Ref.~\cite{Elgood:2020svt}. It is
manifestly not invariant under any electric-magnetic duality transformations.

\section{Restricted, generalized, zeroth laws}
\label{sec-zerothlaws}

Before we derive the Smarr formula and the first law of black hole
thermodynamics we must derive the generalized zeroth laws: the constancy of
the potentials associated to the charges over the event horizon. Our
techniques only allow us to prove them restricted to the bifurcation surface
(hence the name \textit{restricted, generalized zeroth laws}), but this is
sufficient for our purposes. The statements may, in some cases, be extended to
the rest of the horizon using the ideas proposed in
Ref.~\cite{Jacobson:1993vj}

These laws apply to the bifurcation surfaces ($\mathcal{BH}$) of Killing
horizons ($\mathcal{H}$) associated to the Killing vector $k$, which is also
assumed to generate a diffeomorphism that leaves invariant all the fields of
the theory. Thus, $k^{2}\stackrel{\mathcal{H}}{=} 0$,
$k \stackrel{\mathcal{BH}}{=} 0$. In stationary black-hole spacetimes, the
Killing vector whose Killing horizon coincides with the black-hole event
horizon, $k$ is an asymptotically timelike linear combination of the one
generating time translations $t=t^{\mu}\partial_{\mu}$ and those generating
rotations in orthogonal planes $\phi_{n}=\phi_{n}^{\mu}\partial_{\mu}$,

\begin{equation}
  \label{eq:Killingvector}
k=t+\Omega^{n}\phi_{n}\,,  
\end{equation}

\noindent
where the constants $\Omega_{n}$ are the associated angular velocities of the
horizon.

If the $(p+2)$-form field $F$ is invariant under the diffeomorphism generated
by $k$, then we can define the momentum map $p$-form $P_{k}$ satisfying the
momentum map equation (\ref{eq:maxwellmomentummap}) and, assuming that $F$ is
regular on the horizon,

\begin{equation}
dP_{k}= -\imath_{k}F  \stackrel{\mathcal{BH}}{=} 0\,.
\end{equation}

\noindent
Then, using the Hodge decomposition theorem

\begin{equation}
  \label{eq:RGZL}
P_{k} \stackrel{\mathcal{BH}}{=} \Phi^{i}h_{i}+de\,,  
\end{equation}

\noindent
where the $h_{i}$ are harmonic $p$-forms on the bifurcation surface and the
constants $\Phi^{i}$ are going to play the role of potentials associated to
the charges $Q_{i}$ defined in Eq.~(\ref{eq:electriccharges}) now computed by
integration over the bifurcation surface.\footnote{The fact that the charges
  and conjugate potentials that occur in the first law of black-hole mechanics
  and in the Smarr formulae are defined and computed over the horizon cannot
  be overemphasized. Its implications in situations in which the topology of
  the horizon and the topology of spatial infinity are different are dramatic
  \cite{kn:MOPZ}.}

We can also define potentials associated to the magnetic charges. The
invariance of the metric and gauge field under the diffeomorphism generated by
$k$, plus the equations of motion $dG=0$, lead to the existence of a magnetic
momentum map $\tilde{P}_{k}$\footnote{We are going to use the symbol $\doteq$
  for identities that only hold on-shell.}

\begin{equation}
  \label{eq:dualmaxwellmomentummap}
  \delta_{k}G
  \doteq
  -d\imath_{k}G
  =
  0\,,
  \,\,\,\,\,\,
  \Rightarrow
  \,\,\,\,\,\,
  \exists \,\, \tilde{P}_{k}
  \,\,\,\,
  \mid
  \,\,\,\,
  d\tilde{P}_{k}+\imath_{k}G
  \doteq
  0\,.
\end{equation}

In an analogous fashion, the regularity of $G$ over the horizon leads to

\begin{equation}
   \label{eq:RGZLdual}
\tilde{P}_{k} \stackrel{\mathcal{BH}}{=} \Phi_{m}\tilde{h}^{m}+de\,,  
\end{equation}

\noindent
where the $\tilde{h}^{m}$ are harmonic $\tilde{p}$-forms on the bifurcation
surface and the constants $\Phi_{m}$ are going to play the role of potentials
associated to the magnetic charges $P^{m}$ defined in
Eq.~(\ref{eq:magneticcharges}), now computed by integration over the
bifurcation surface.

Observe that the same reasoning can be applied to the Lorentz momentum map
equation, obtaining

\begin{equation}
  \mathcal{D}P_{k}{}^{ab}
  \stackrel{\mathcal{BH}}{=}
  0\,,
\end{equation}

\noindent
which implies that $P_{k}{}^{ab}$ can be expanded as a linear combination with
constant coefficients of covariantly constant antisymmetric Lorentz
tensors. It is a well-known result that 

\begin{equation}
  \label{eq:Pkab=knab}
P_{k}{}^{ab}
\stackrel{\mathcal{BH}}{=}
\kappa n^{ab}\,,
\end{equation}

\noindent
where $\kappa$ is the surface gravity (constant over the whole event horizon,
according to the standard zeroth law) and $n^{ab}$ is the binormal to the
horizon with the normalization $n^{ab}n_{ab}=-2$. Clearly, $n^{ab}$ is
covariantly constant over the bifurcation surface and $\kappa$ can be
interpreted as the ``potential'' associated to the Lorentz charge, which is,
essentially, the area of the (spatial sections of the) horizon.

\section{Smarr formulae}
\label{sec-Smarrformulae}

Smarr formulae for stationary black-hole solutions \cite{Smarr:1972kt} can be
systematically obtained using Komar integrals
\cite{Komar:1958wp,Kastor:2008xb,Kastor:2010gq}.  Wald's formalism, in its
turn, can be used to construct the $(d-2)$-form integrands of Komar integrals,
that we are going to call \textit{Komar charges} as explained in
Refs.~\cite{Liberati:2015xcp,Ortin:2021ade,Mitsios:2021zrn,Meessen:2022hcg}
(see also \cite{Jacobson:2018ahi}).

The main observation is that, on-shell and for a Killing vector $k$ that
generates a symmetry of the whole field configuration, the only non-vanishing
contribution to $\mathbf{J}[k]$ is $\imath_{k}\mathbf{L}$

\begin{equation}
  \label{eq:JiL}
\mathbf{J}[k] \doteq\imath_{k}\mathbf{L}\,.  
\end{equation}

\noindent
Furthermore, under the same conditions,

\begin{equation}
0=-\delta_{k}\mathbf{L} \doteq\pounds_{k}\mathbf{L}= d\imath_{k}\mathbf{L}\,,  
\end{equation}

\noindent
which implies the local existence of the $(d-2)$-form $\omega_{k}$ 

\begin{equation}
d \omega_{k} \doteq \imath_{k}\mathbf{L}\,.  
\end{equation}

\noindent
Since we have proven that $\mathbf{J}[\xi] = d\mathbf{Q}[\xi]$,
Eq.~(\ref{eq:JiL}) leads to the identity

\begin{equation}
  \label{eq:Komaridentity}
  d\mathbf{K}[k]\doteq 0\,,
\end{equation}

\noindent
for the \textit{Komar charge} $(d-2)$-form $ \mathbf{K}[k]$ defined by 

\begin{equation}
  \label{eq:Komarchargedef}
  \mathbf{K}[k]
  \equiv
-\left(\mathbf{Q}[k]-\omega_{k}\right)\,.
\end{equation}

Smarr formulae for stationary black holes are obtained by integrating
Eq.~(\ref{eq:Komaridentity}) on hypersurfaces $\Sigma$ with boundaries at a
spatial section of the event horizon $\partial \Sigma_{h}$ (usually, the
bifurcation surface $\mathcal{BH}$) and at spatial infinity
$\partial\Sigma_{\infty}$. Applying Stokes' theorem to that integral one gets

\begin{equation}
  \label{eq:KomarandSmarr}
  \int_{\partial\Sigma_{\infty}} \mathbf{K}[k]
  =
  \int_{\mathcal{BH}}\mathbf{K}[k]\,,
\end{equation}

\noindent
and performing the integrals one arrives at the Smarr formula.

In order to apply this algorithm we must first construct the Komar charge
$\mathbf{K}[k]$ finding $\omega_{k}$. This can be done for general
configurations using the techniques of Ref.~\cite{Mitsios:2021zrn}. The trace
of the Einstein equation (\ref{eq:Ea}) can be written in terms of the
Lagrangian as follows:

\begin{equation}
  e^{a}\wedge \mathbf{E}_{a}
  =
  (-1)^{d-1}(d-2)
  \left\{\mathbf{L} -(-1)^{d(p+1)}\frac{(p+1)}{(d-2)}F\wedge G\right\}\,,
\end{equation}

\noindent
which implies that the on-shell Lagrangian takes the value

\begin{equation}
  \mathbf{L}
  \doteq
  (-1)^{d(p+1)}\frac{(p+1)}{(d-2)}F\wedge G\,.  
\end{equation}

\noindent
Next, using the momentum map equations (\ref{eq:maxwellmomentummap}) and
(\ref{eq:dualmaxwellmomentummap})

\begin{equation}
  \imath_{k}\mathbf{L}
  \doteq
  -(-1)^{d(p+1)}\frac{(p+1)}{(d-2)}\left[dP_{k}\wedge G
    +(-1)^{p(d-1)}d\tilde{P}_{k}\wedge F\right]\,.  
\end{equation}

\noindent
and integrating by parts and using the equation of motion and Bianchi
identity, we arrive at

\begin{equation}
  \omega_{k}
  =
  -(-1)^{d(p+1)}\frac{(p+1)}{(d-2)}\left[P_{k}\wedge G
    +(-1)^{p(d-1)}\tilde{P}_{k}\wedge F\right]\,.
\end{equation}

The Komar charge $(d-2)$-form is, then, given by 

\begin{equation}
  \label{eq:Komarcharge}
  \begin{aligned}
    \mathbf{K}[k]
    & =
    \frac{1}{16\pi G_{N}^{(d)}} \left\{(-1)^{d-1}\star
      (e^{a}\wedge e^{b})P_{k\, ab}
    \right.
    \\
    & \\
    & \hspace{.5cm}
    \left.
      +\frac{(-1)^{d(p+1)}}{d-2}\left[(\tilde{p}+1)P_{k}\wedge G
      -(-1)^{d}\sigma^{2}(p+1)\tilde{P}_{k}\wedge F\right]
    \right\}\,.
  \end{aligned}
\end{equation}

When $p=\tilde{p}$ (so $d$ is even), defining the vector of momentum maps

\begin{equation}
  \mathcal{P}_{k}
  \equiv
  \left(
    \begin{array}{c}
     P_{k} \\ \tilde{P}_{k} \\ 
    \end{array}
  \right)\,,
\end{equation}

\noindent
which transforms as $\mathcal{F}$ under electric-magnetic duality
because it satisfies the equation

\begin{equation}
  d  \mathcal{P}_{k} +\imath_{k}\mathcal{F}=0\,,
\end{equation}

\noindent
$\mathbf{K}[k]$ can be rewritten in the manifestly duality-symmetric form

\begin{equation}
  \label{eq:Komarchargeptildep}
  \begin{aligned}
    \mathbf{K}[k]
    & =
    \frac{1}{16\pi G_{N}^{(d)}} \left\{(-1)^{d-1}\star
      (e^{a}\wedge e^{b})P_{k\, ab}
      +\frac{(p+1)}{(d-2)}\mathcal{P}^{T}_{k}\wedge \Omega\mathcal{F}
    \right\}\,.
  \end{aligned}
\end{equation}

We now plug the Komar charge Eq.~(\ref{eq:Komarcharge}) in the integrals of
Eq.~(\ref{eq:KomarandSmarr}). For asymptotically-flat black holes, only the
gravitational term in the first line contributes to the integral over spatial
infinity since the products of potentials and gauge fields fall off too fast
approaching infinity if we impose adequate boundary conditions. Using also the
restricted generalized zeroth laws for the momentum maps Eqs.~(\ref{eq:RGZL})
and (\ref{eq:RGZLdual}), we get

\begin{equation}
  \begin{aligned}
\frac{1}{16\pi G_{N}^{(d)}}\int_{\partial\Sigma_{\infty}} (-1)^{d-1}\star
      (e^{a}\wedge e^{b})P_{k\, ab}
  & =
\frac{1}{16\pi G_{N}^{(d)}}\int_{\mathcal{BH}} (-1)^{d-1}\star
(e^{a}\wedge e^{b})P_{k\, ab}
\\
& \\
& \hspace{.5cm}
      +\frac{1}{(d-2)}\left[(\tilde{p}+1)\Phi^{i}Q_{i}
 +(-1)^{d}(p+1)\sigma^{2}\tilde{\Phi}_{m}P^{m}\right]\,.
    \end{aligned}
\end{equation}

For the Killing vector Eq.~(\ref{eq:Killingvector}), the integral in the
left-hand side of this equation gives

\begin{equation}
\frac{1}{16\pi G_{N}^{(d)}}\int_{\partial\Sigma_{\infty}} (-1)^{d-1}\star
      (e^{a}\wedge e^{b})P_{k\, ab}
      =
        \frac{(d-3)}{(d-2)}\left(M -\Omega^{n}J_{n}\right)\,,  
\end{equation}

\noindent
where $M$ is the mass and $J_{n}$ are the components of the angular momentum.
Furthermore, using Eq.~(\ref{eq:Pkab=knab}), the integral in the right-hand
side gives

\begin{equation}
\frac{1}{16\pi G_{N}^{(d)}}\int_{\mathcal{BH}} (-1)^{d-1}\star
(e^{a}\wedge e^{b})P_{k\, ab}
     =
     -\frac{1}{16\pi G_{N}^{(d)}}\int_{\mathcal{BH}} dA n^{ab} P_{k\, ab}
     =
     \frac{\kappa A}{8\pi G_{N}^{(d)}}
     =
        TS\,,
\end{equation}

\noindent
where $T$ is  the Hawking temperature and $S$ is the Bekenstein-Hawking entropy.

Thus, we get the  Smarr equation

\begin{equation}
  \label{eq:Smarr}
  M 
  =
  \frac{(d-2)}{(d-3)}TS
  +\Omega^{n}J_{n}
  +\frac{(\tilde{p}+1)}{(d-3)}\Phi^{i}Q_{i}
 +(-1)^{d}\sigma^{2}\frac{(p+1)}{(d-3)}\tilde{\Phi}_{m}P^{m}\,.
\end{equation}

For $p=\tilde{p}$ this formula takes the manifestly electric-magnetic duality
invariant form 

\begin{equation}
  \label{eq:Smarrptildep}
  M 
  =
  \frac{(d-2)}{(d-3)}TS
  +\Omega^{n}J_{n}
  +\frac{(p+1)}{(d-3)} \hat{\Phi}^{T\,i}\wedge \Omega\mathcal{Q}_{i}\,,
\end{equation}

\noindent
where $\mathcal{Q}_{i}$ is the charge vector defined in
Eq.~(\ref{eq:chargevector}) and $\hat{\Phi}^{i}$ is the vector of potentials

\begin{equation}
  \hat{\Phi}^{i}
  \equiv
  \left(
    \begin{array}{c}
     \tilde{\Phi}^{i} \\ \Phi^{i} \\ 
    \end{array}
    \right)\,,
\end{equation}

\noindent
so that 

\begin{equation}
  \hat{\Phi}^{T\,i}\wedge \Omega\mathcal{Q}_{i}
  =
\Phi^{i}Q_{i}
 +\sigma^{2}\tilde{\Phi}_{i}P^{i}\,.
\end{equation}

\section{First law}
\label{sec-firstlaw}

We are going to review the derivation of the first law in full detail,
improving the derivations made in
Refs.~\cite{Elgood:2020svt,Elgood:2020mdx,Elgood:2020nls,Mitsios:2021zrn,Meessen:2022hcg}
and showing where and how the variation of the magnetic charges, missed in
those works, arise.

Following Refs.~\cite{Lee:1990nz,Wald:1993nt,Iyer:1994ys}, and denoting by
$\varphi$ all the fields of the theory, we define the symplectic $(d-1)$-form

\begin{equation}
  \omega(\varphi,\delta_{1}\varphi,\delta_{2}\varphi)
  \equiv
  \delta_{1}\mathbf{\Theta}(\varphi,\delta_{2}\varphi)
-\delta_{2}\mathbf{\Theta}(\varphi,\delta_{1}\varphi)\,,
\end{equation}

\noindent
and we choose $\delta_{1}\varphi=\delta\varphi$, variations which satisfy the
linearized equations of motion but which are, otherwise, arbitrary, and
$\delta_{2}\varphi=\delta_{\xi}\varphi$, the transformations under
diffeomorphisms that we have defined in
Section~\ref{sec-NoetherWaldcharge}. On-shell
$\mathbf{\Theta}=\mathbf{\Theta}'$ and using the definitions of
$\mathbf{J}[\xi]$ Eq.~(\ref{eq:Jxidef}) and $\delta_{\xi}$
Eq.~(\ref{eq:deltaxidef})

\begin{equation}
  \begin{aligned}
    \omega(\varphi,\delta\varphi,\delta_{\xi}\varphi)
    & \doteq
    \delta\mathbf{\Theta}'(\varphi,\delta_{\xi}\varphi)
    -\delta_{\xi}\mathbf{\Theta}'(\varphi,\delta\varphi)
    \\
    & \\
    & =
    \delta\left(\mathbf{J}[\xi]-\imath_{\xi}\mathbf{L}\right)
    -\left(-\pounds_{\xi}+\delta_{\Lambda_{\xi}}\right)
    \mathbf{\Theta}'(\varphi,\delta\varphi)
    \\
    & \\
    & =
    \delta\mathbf{J}[\xi]
    -\imath_{\xi}\delta\mathbf{L}
    +\left(\imath_{\xi}d+d\imath_{\xi}-\delta_{\Lambda_{\xi}}\right)
    \mathbf{\Theta}'(\varphi,\delta\varphi)
    \\
    & \\
    & =
    \delta d\mathbf{Q}[\xi]
    -\imath_{\xi}\left(\mathbf{E}_{\varphi}\wedge \delta\varphi
      +d \mathbf{\Theta}(\varphi,\delta\varphi)\right)
    +\left(\imath_{\xi}d+d\imath_{\xi}-\delta_{\Lambda_{\xi}}\right)
    \mathbf{\Theta}'(\varphi,\delta\varphi)
    \\
    & \\
    & \doteq
    d\left[\delta \mathbf{Q}[\xi]
    +\imath_{\xi} \mathbf{\Theta}'(\varphi,\delta\varphi)\right]
    -\delta_{\Lambda_{\xi}}\mathbf{\Theta}'(\varphi,\delta\varphi)\,.    
  \end{aligned}
\end{equation}

This result differs from the standard one by the last term, which does not
look like a total derivative. Let us study it in more detail in the theory at
hand:

\begin{equation}
  \begin{aligned}
    \delta_{\Lambda_{\xi}}\mathbf{\Theta}'(\varphi,\delta\varphi)
    & \doteq
    \delta_{\Lambda_{\xi}}\left\{-\star (e^{a}\wedge e^{b})\wedge \delta \omega_{ab}
      +G\wedge \delta A\right\}
    \\
    & \\
    & =
    -\delta_{\sigma_{\xi}}\left\{\star (e^{a}\wedge e^{b}) \right\}\wedge \delta\omega_{ab}
    -\star (e^{a}\wedge e^{b})\wedge \delta_{\sigma_{\xi}}\delta \omega_{ab}
    +G\wedge \delta_{\chi_{\xi}}\delta A\,,
  \end{aligned}
\end{equation}

\noindent
since $G$ is gauge invariant. Now, let us consider the last term. By
definition, and taking into account that the parameter of the compensating
gauge transformation depends on the field on which the transformation acts

\begin{equation}
    \delta_{\chi_{\xi}}\delta A
     =
    \delta_{\chi_{\xi}}(A'-A)
    =
    d\chi(\xi,A') - d\chi(\xi,A)
    =
    d\delta \chi_{\xi}\,.
\end{equation}

\noindent
By the same token

\begin{equation}
  \delta_{\sigma_{\xi}}\delta \omega_{ab}
  =
  \mathcal{D}\delta \sigma_{\xi} -2\delta \omega_{[a|}{}^{c}\sigma_{\xi\, c|b]}\,,
\end{equation}

\noindent
while the first term transforms in the standard fashion

\begin{equation}
  \delta_{\sigma_{\xi}}\left\{\star (e^{a}\wedge e^{b}) \right\}
  =
  2\sigma_{\xi}{}^{[a|}{}_{c}\star (e^{c}\wedge e^{|b]})\,.
\end{equation}

Combining these results, integrating by parts and using the equation of motion
$dG=0$ we arrive at another total derivative

\begin{equation}
  \begin{aligned}
    \delta_{\Lambda_{\xi}}\mathbf{\Theta}'(\varphi,\delta\varphi)
    & \doteq
    -\left\{\star (e^{a}\wedge e^{b}) \right\}\wedge \mathcal{D}\delta\sigma_{\xi\, ab}
    +G\wedge  d\delta \chi_{\xi}
    \\
    & \\
    & \doteq
    d\left\{    (-1)^{d-1}\star (e^{a}\wedge e^{b})\delta \sigma_{\xi\, ab}
    +(-1)^{\tilde{p}}G\wedge  \delta \chi_{\xi}
 \right\}\,,
  \end{aligned}
\end{equation}

\noindent
which allows us to rewrite the complete symplectic $(d-1)$-form as the total
derivative of a $(d-2)$-form that we will denote by
$\mathbf{\Omega}(\varphi,\delta\varphi,\delta_{\xi}\varphi)$, which is defined up
to total derivatives

\begin{subequations}
  \begin{align}
    \omega(\varphi,\delta\varphi,\delta_{\xi}\varphi)
    & \doteq
      -d\mathbf{\Omega}(\varphi,\delta\varphi,\delta_{\xi}\varphi)\,,
    \\
    & \nonumber \\
    \mathbf{\Omega}(\varphi,\delta\varphi,\delta_{\xi}\varphi)
    & \equiv
-\delta \mathbf{Q}[\xi]
      -\imath_{\xi} \mathbf{\Theta}'(\varphi,\delta\varphi)
    \nonumber \\
    & \nonumber \\
    & \hspace{.5cm}
  +\frac{1}{16\pi G_{N}^{(d)}}\left\{(-1)^{d-1}\star (e^{a}\wedge e^{b})\delta \sigma_{\xi\, ab}
  +(-1)^{\tilde{p}}G\wedge  \delta \chi_{\xi}\right\}\,.  
  \end{align}
\end{subequations}

Plugging into $\mathbf{\Omega}(\varphi,\delta\varphi,\delta_{\xi}\varphi)$ the
expressions we have obtained for $\mathbf{Q}[\xi]$ and
$\mathbf{\Theta}'(\varphi,\delta\varphi)$ and operating, we can put 
$\mathbf{\Omega}(\varphi,\delta\varphi,\delta_{\xi}\varphi)$ in this form:

\begin{equation}
  \begin{aligned}
    \mathbf{\Omega}(\varphi,\delta\varphi,\delta_{\xi}\varphi)
    & =
    (-1)^{d-1}\delta \star (e^{a}\wedge e^{b})\wedge P_{\xi\, ab}
    +\imath_{\xi} \star (e^{a}\wedge e^{b})\wedge \delta \omega_{ab}
    \\
    & \\
    & \hspace{.5cm}
    +(-1)^{d(p+1)}P_{\xi}\wedge \delta G -\imath_{\xi}G \wedge \delta A\,,
  \end{aligned}
\end{equation}

\noindent
again, up to total derivatives. We are going to profit from this freedom to
rewrite this charge as follows:

\begin{equation}
  \begin{aligned}
    \mathbf{\Omega}(\varphi,\delta\varphi,\delta_{\xi}\varphi) & =
    \delta\left[(-1)^{d-1}\star (e^{a}\wedge e^{b})\wedge P_{\xi\, ab}\right]
    -(-1)^{d-1}\star (e^{a}\wedge e^{b})\wedge \delta P_{\xi\, ab}
    \\
    & \\
    & \hspace{.5cm}
    +\imath_{\xi} \star (e^{a}\wedge e^{b})\wedge \delta \omega_{ab}
    +(-1)^{d(p+1)}P_{\xi}\wedge \delta G
    -\left(\imath_{\xi}G+d\tilde{P}_{\xi}\right) \wedge \delta A
    \\
    & \\
    & \hspace{.5cm}
    +(-1)^{\tilde{p}+1}\tilde{P}_{\xi}\wedge \delta F\,.
  \end{aligned}
\end{equation}

Now, when $\xi=k$, since $\delta_{k}\varphi=0$ implies
$\omega(\varphi,\delta\varphi,\delta_{k}\varphi)=0$, we have the identity

\begin{equation}
  \label{eq:Widentity}
  d\mathbf{\Omega}(\varphi,\delta\varphi,\delta_{k}\varphi)
  \doteq
  0\,,
\end{equation}

\noindent
where, upon use of the definition of the dual momentum map
Eq.~(\ref{eq:dualmaxwellmomentummap})
$\mathbf{\Omega}(\varphi,\delta\varphi,\delta_{k}\varphi)$ takes the final form

\begin{equation}
  \label{eq:Wmassaged}
  \begin{aligned}
    \mathbf{\Omega}(\varphi,\delta\varphi,\delta_{k}\varphi) & =
 \delta\left[   (-1)^{d-1}\star (e^{a}\wedge e^{b})\wedge P_{k\, ab}\right]
    -(-1)^{d-1}\star (e^{a}\wedge e^{b})\wedge \delta P_{k\, ab}
    \\
    & \\
    & \hspace{.5cm}
    +\imath_{k} \star (e^{a}\wedge e^{b})\wedge \delta \omega_{ab}
    +(-1)^{d(p+1)}\left[P_{k}\wedge \delta G
    +(-1)^{d}\sigma^{2}\tilde{P}_{k}\wedge \delta F\right]\,.
  \end{aligned}
\end{equation}

To proceed, we integrate the identity Eq.~(\ref{eq:Widentity}) over the same
hypersurface over which we integrated the analogous identity involving the
Komar charge $\mathbf{K}[k]$ in the previous section. Using Stokes' theorem

\begin{equation}
\int_{\partial\Sigma_{\infty}} \mathbf{\Omega}(\varphi,\delta\varphi,\delta_{k}\varphi)
  =
  \int_{\mathcal{BH}}\mathbf{\Omega}(\varphi,\delta\varphi,\delta_{k}\varphi)\,.
\end{equation}

For the Killing vector Eq.~(\ref{eq:Killingvector}) the integral at spatial
infinity can be shown to give \cite{Iyer:1994ys,Barnich:2005kq}\footnote{As in
  the calculation of the Smarr formula, the additional terms that we have
  found will not contribute at infinity if we impose suitable boundary
  conditions to the fields and their variations.}

\begin{equation}
  \int_{\partial\Sigma_{\infty}}
  \mathbf{\Omega}(\varphi,\delta\varphi,\delta_{k}\varphi)
  =
 \delta M -\Omega_{n}\delta J^{n}\,. 
\end{equation}

When evaluating the integral over the bifurcation surface, we can use the
reasoning in Ref.~\cite{Iyer:1994ys} to show that the second term in
Eq.~(\ref{eq:Wmassaged}) does not contribute and that the first gives, simply
$\kappa\delta A/(8\pi G_{N}^{(d)})$. The third simply vanishes on the
bifurcation surface. Using these results, the restricted, generalized, zeroth
laws Eqs.~(\ref{eq:RGZL}) and (\ref{eq:RGZLdual}) and the definitions of
electric and magnetic charges Eqs.~(\ref{eq:electriccharges}) and
(\ref{eq:magneticcharges}), the integral over the bifurcation surface gives

\begin{equation}
    \int_{\mathcal{BH}} \mathbf{\Omega}(\varphi,\delta\varphi,\delta_{k}\varphi)
    =
        \frac{\kappa\delta A}{8\pi G_{N}^{(d)}}
    +(-1)^{d(p-1)}\left[\Phi^{i}\delta Q_{i}
      +(-1)^{d}\sigma^{2}\tilde{\Phi}_{m}\delta P^{m}\right]\,,
\end{equation}

\noindent
and we arrive at the first law

\begin{equation}
  \label{eq:firstlaw}
  \delta M
  =
  \frac{\kappa\delta A}{8\pi G_{N}^{(d)}}  +\Omega_{n}\delta J^{n}
  +(-1)^{d(p-1)}\left[\Phi^{i}\delta Q_{i}
    +(-1)^{d}\sigma^{2}\tilde{\Phi}_{m}\delta P^{m}\right]\,,
\end{equation}

\noindent
which, for the $\tilde{p}=p$ case takes the manifestly electric-magnetic
duality-invariant form 

\begin{equation}
  \label{eq:firstlawptildep}
  \delta M
  =
  \frac{\kappa\delta A}{8\pi G_{N}^{(d)}}  +\Omega_{n}\delta J^{n}
  +\Phi^{i}\Omega \delta \mathcal{Q}_{i}\,.
\end{equation}

\section{Discussion}
\label{sec-discussion}

In this paper we have studied how to deal with magnetic charges in a
$d$-dimensional generalization of the Einstein-Maxwell theory with
$(p+1)$-form potentials. Our main results are

\begin{enumerate}
\item The Komar charge Eqs.~(\ref{eq:Komarcharge}), which, for $p=\tilde{p}$,
  takes the manifestly electric-magnetic duality-invariant form
  Eq.~(\ref{eq:Komarchargeptildep}).
\item The Smarr formula Eq.~(\ref{eq:Smarr}), which, again, takes the
  manifestly electric-magnetic duality-invariant form Eq.~(\ref{eq:Smarrptildep}) in the
  $p=\tilde{p}$ case.
\item The first law Eq.~(\ref{eq:firstlaw}) and the manifestly
  electric-magnetic duality-invariant form Eq.~(\ref{eq:firstlawptildep}) that
  it takes when $p=\tilde{p}$.
\end{enumerate}

We have assumed in the derivation of these results the asymptotic flatness of
the solutions. Thus, they are valid for black holes, black rings and their
generalizations, but, in order to apply them to infinite, planar, $p$-branes,
a few, simple, modifications would be necessary to replace mass by tension and
charges by charge densities removing the infinite volume factors. Wrapping
these branes on compact dimensions would introduce additional effects (KK and
winding modes) that need to be studied separately.\footnote{Work in this
  direction is in progress  \cite{kn:MOPZ}.}

Furthermore, observe that the Smarr formulae and first laws obtained are
generic: a particular solution may not be able to carry the electric, the
magnetic or either charge. For instance, a black hole in 6 dimensions in a
theory with a 2-form will not be able to carry electric nor magnetic charge
with respect to the 2-form. In 5 dimensions, a black hole can carry the
electric charge of a 1-form potential but not the magnetic charge (electric
with respect to a 2-form potential), while a black ring can, in principle,
carry the opposite.

In the $\tilde{p}=p$ cases, black $p$-branes can carry electric and magnetic
charges of the same $(p+1)$ potential and, as it is well known, since
electric-magnetic duality leaves invariant the metric, all their geometric
properties including their surface gravity and area are also duality
invariant. Thus, the first law of their dynamics should also be invariant. Our
results show that this is, indeed, the case.

As mentioned in the Introduction, the first law also has a term proportional
to the variation of the moduli where the proportionality constants are the
scalar charges, for which no good definition as conserved charges has ever
been given \cite{Gibbons:1996af}. Here we have avoided this problem by
studying a theory with no scalar fields, but this is a problem that has to be
confronted and understood and we plan to do so in future work.

\section*{Acknowledgments}

D.P.~would like to thank Profs.~Roberto Emparan and David Mateos and
T.O.~would like to thank Prof.~Glenn Barnich for useful and friendly
conversations. This work has been supported in part by the MCIU, AEI, FEDER
(UE) grant PGC2018-095205-B-I00 and by the Spanish Research Agency (Agencia
Estatal de Investigaci\'on) through the grant IFT Centro de Excelencia Severo
Ochoa CEX2020-001007-S.  The work of DP is supported by a ``Campus de
Excelencia Internacional UAM/CSIC'' FPI pre-doctoral grant. TO wishes to thank
M.M.~Fern\'andez for her permanent support.

\appendix


\end{document}